     \documentstyle[prl,aps,epsfig,epsf]{revtex} 
     \newcommand {\tpr} {t^\prime} 
     
\begin{document}
     \draft
     \title{\bf{Low density ferromagnetism in Extended Hubbard (
     t-t$^\prime$-$U$ ) Model}} 
     \author{P. A. Sreeram, Suresh G. Mishra} 
     \address{Institute of Physics, Bhubaneswar 751005, India} 
     \maketitle 
     \begin{abstract}

     We study the existence of ferromagnetism in one dimensional Hubbard  
     Model with on site interaction and nearest and next nearest neighbor
     hopping. Using the Hubbard I approximation, a self consistent
     equation for $ < n_\sigma > $, the particle density with spin
     $\sigma$ is obtained. We find that ferromagnetism exists over a
     wide range of values of the next nearest neighbor hopping 
     element $ t^\prime$ when on-site interaction $ U $ is large. The
     phase diagram as a function of $ \tpr / t $ and the electron
     density $ < n > $ has been obtained by numerically solving the
     self consistent equation. It is found that the maximum density
     at which ferromagnetism can occur, peaks around
     $ \tpr / t \approx -0.53 $. We show that this phenomenon is a
     result of the peculiar way in which the $t^\prime$ term modifies
     the density of states.  

     \end{abstract} 


     For last thirty years various techniques have been used in the
     study of the existence of ferromagnetism within the Hubbard
     Model. Within the t-matrix approximation Kanamori \cite{1} had shown,
     that, for a given on-site repulsion $ U $ there exists a critical
     density $\rho_c$ below which no ferromagnetism can exist. Mean
     Field theory, however, allows the existence of ferromagnetism as
     long as the Stoner Criterion $ U \rho(\epsilon_f) > 1$ is
     satisfied, where $\rho(\epsilon_f)$ is the density of states at
     Fermi energy, $\epsilon_f$. In his first paper (henceforth, we shall
     refer to this as Hubbard-I)\cite{2}, Hubbard himself
     had asserted that, if the density of states is peaked away from
     the center of the band, then the ferromagnetic phase can exist
     for low density of electrons. The Lieb-Mattis theorem\cite{3}, however,
     precludes any possibility of existence of ferromagnetism in the
     1-dimensional Hubbard Model for any arbitrary filling. 

     On the other hand, in certain limiting cases, e.g. the Nagaoka
     point \cite{4,5} , at $ U= \infty$ and a one hole in an otherwise 
     half filled
     band it is possible to show the existence of ferromagnetism.
     Another example where saturated ferromagnetism has been shown to
     exist in Hubbard Model is the case of Flat Band Ferromagnetism
     \cite{6}. In this mechanism, the ground state of the 
      Hamiltonian
     H = H$_0$ + H$_{int}$ is at the same time the ground state of H$_0$
     and H$_{int}$ separately. This model can be mapped on to a degenerate
     band Hubbard Model.

     Recently M\"uller-Hartman \cite{7} has shown that it is possible to get
     ferromagnetism in 1-D Hubbard Model at low electron densities,
     if one introduces next-nearest neighbor hopping 
     ($t^\prime$), besides the nearest neighbor hopping matrix 
     element $ t $, for
     the condition $ t^\prime / t < -0.25 $. This work has been extended
     by Pieri et. al. \cite{8} using the G\"utzwiller approach. In the
     present work,
     we study the $ t  -  t^\prime  -  U $ model using the Hubbard I
     approximation. We find M\"uller-Hartman result reminiscent of
     the comment made in the Hubbard first paper. The introduction of
     the $t^{\prime}$ term brings more weightage to the lower edge of
     the density of states, moreover it also takes care of the limitation
     concerning the Lieb Mattis theorem. We now give some detail of
     our work.  

     The model under consideration is given by the Hamiltonian,
     \begin{equation}
     H = -t \sum_{i \sigma} (C_{i \sigma}^\dagger C_{i+1 \sigma} +
     C_{i+1 \sigma}^\dagger C_i )- t^\prime \sum_{i} 
     ( C_{i \sigma}^\dagger C_{i+2 \sigma} + C_{i+2 \sigma}^\dagger 
     C_{i \sigma}) + U \sum_{i} n_{i \uparrow} n_{i \downarrow}
     \end{equation}
     where C$_{i \sigma}^\dagger$ (C$_{i \sigma} $) is the creation
     (annihilation) operator of an electron with spin $\sigma$ at the
     site $ i $, $n_{i\sigma}$ = $C_{i\sigma}^\dagger C_{i\sigma}$ and
     t,$\tpr$ and U have the usual meaning, as defined earlier. 
     The noninteracting Hamiltonian can be diagonalized
     in the momentum space giving,
     \begin{equation}
     H_0(k) = \sum_{k \sigma} \epsilon_k C_{k \sigma}^\dagger 
     C_{k\sigma}
     \end{equation}
     where $\epsilon_k$ takes the form,
     \begin{equation}
     \epsilon_k = -2t cos(k)-2t^\prime cos(2k), 
     \end{equation}
     \begin{center} 
      \epsfig{file=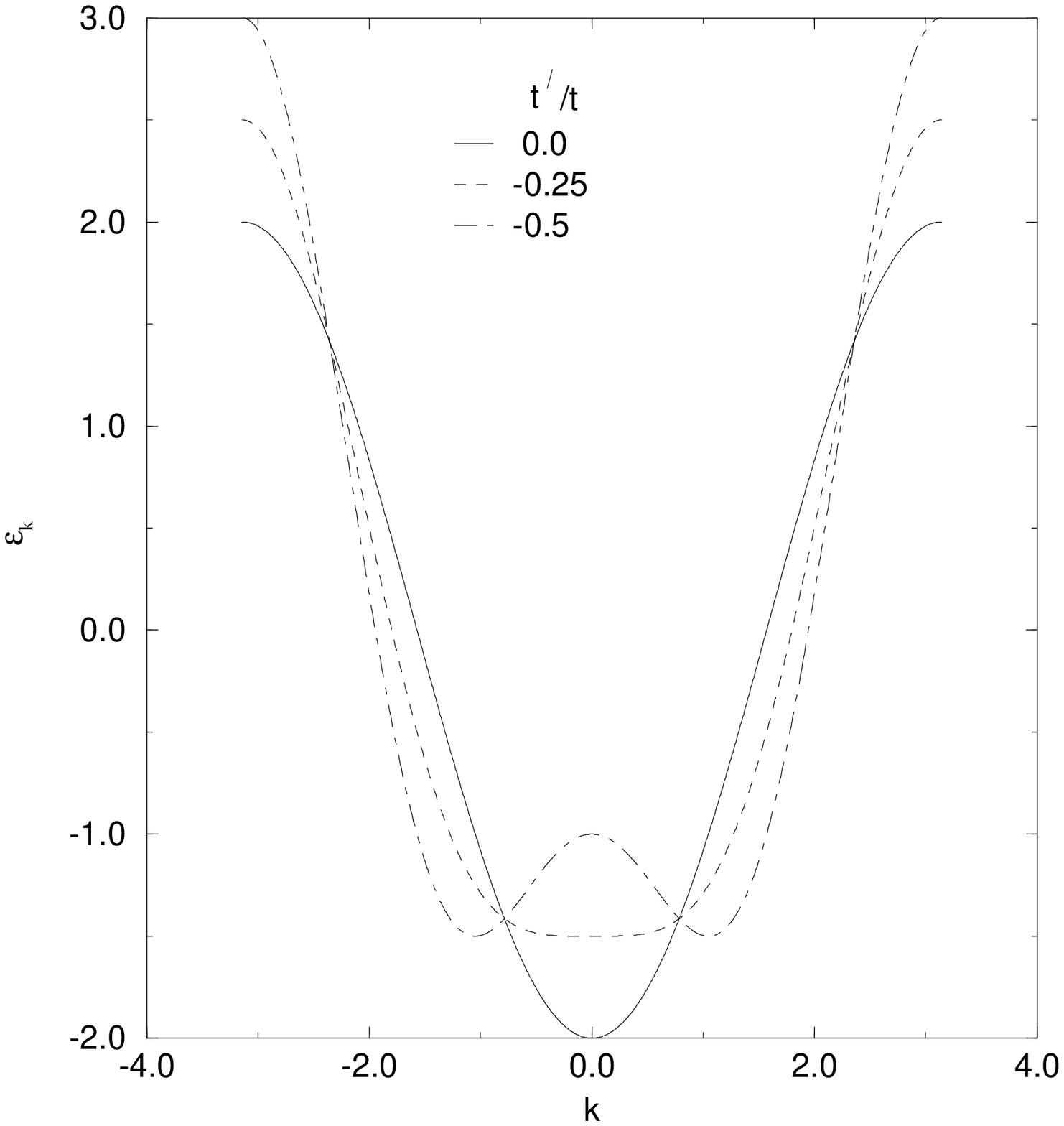,width=9cm,height=7cm}
      \vspace{-1.0cm}
      \begin{figure}[p]
      \caption{Free electron dispersion curve for 
      $\tpr$/t = 0.0,-0.25 and -0.5.
      \label{disp}} 
      \vspace{0.5cm}
      \end{figure}
      \end{center}
     for one dimension.
     The dispersion is parabolic at low $ k $ and develops a double
     well structure as the value of $\tpr /t $ reduced below - 0.25
     (Fig. 1). It is shown below that the existence of ferromagnetism in
     the system is closely related to this particular structure of
     the dispersion.  

     The Greens Function, $G_{\sigma}(k,\omega)=<<C_k;C_k^\dagger>>$,
     for the Hamiltonian given in Eq. 1, can be calculated using the
     Memory Function Technique \cite{9}; it is given by, 
     \begin{equation}
     G_{\sigma}(k,\omega)=[ \omega - \epsilon_k - U <n_{-\sigma}> 
     (1+\frac {\hat U} {\omega - \hat U})+i \eta]^{-1}
     \end{equation}
     upto the first order in the memory kernel, 
     where, $\hat U = U (1- <n_{-\sigma}>)$, and $\epsilon_k$ is
     given by Eq. 3. The number density $<n_{\sigma}>$ is related to
     the Greens function by, 
     \begin{equation}
     <n_{\sigma}> = \frac{1}{2\pi}\int_{-\infty}^{\mu}\sum_k 
     Im G_{\sigma} (k,\omega).
     \end{equation}
     We thus have a self consistent equation for $<n_{\sigma}>$. 
     The phase diagrams obtained by solving Eq. 5 are 
     \begin{center}
      \epsfig{file=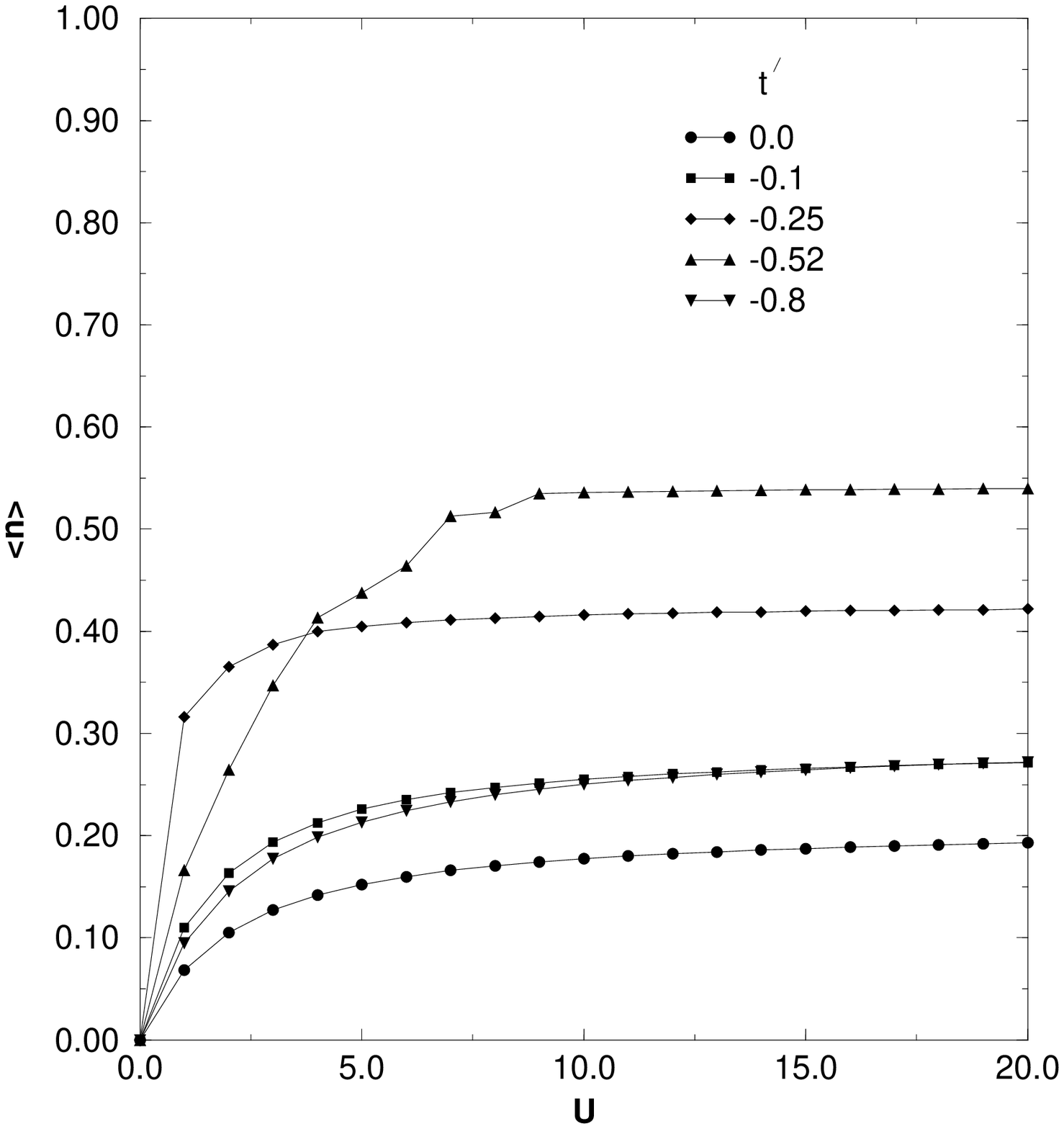,width=9cm,height=6cm}
      \vspace{-1.0cm}
      \begin{figure}[p]
      \caption{Phase diagram in the $U$-$<n>$ plane for $\tpr$/t = 0.0,
      -0.1, -0.25, -0.52 and -0.8. The ferromagnetic phase exists 
      below the curve corresponding to each value of $\tpr$/t.
      \label{Uphase1}}
      \vspace{0.5cm}
      \end{figure}
      \end{center}
     shown in Figs. 2 and 3.
     In Fig. 2 we show the phase diagram as a function of electron
     density $<n>$ and the on-site repulsion $U$, for various values of
     the next nearest neighbor hoping parameter t$^\prime$. From
     the figure, it is obvious that for very low value of $U$, the
     region of ferromagnetic phase is very small. As $U$ increases,
     the critical density of electrons $<n>_c$ up to which
     ferromagnetism can exist increases, but this critical density
     saturates beyond a certain value of $U$. Thus, for any given value
     of $\tpr /t $, there exists a maximum value of $<n>_c$ beyond
     which ferromagnetism cannot exist, even if $U$ is increased
     indefinitely.
     The dependence on $\tpr/t$ is also clear from this figure. For
     large U ( when $<n>_c$ has saturated ), there exists a optimum
     value of $\tpr/t$ ($\approx$ -0.52) which gives maximum $<n>_c$.
     This becomes very clear from the phase diagram shown in
     Fig-3. However, at low values of U, this optimum value of $\tpr/t$ 
     shifts towards $\approx$ -0.25.
     
     Fig.3 shows the phase diagram in the $<n>$-$ t^\prime$ plane for
     $U$=10. 
     The value of U has been such that the critical value of $<n>_c$ 
     has reached the saturated value for large U. Thus this phase diagram
     is a reflection of the large U behaviour. We have chosen this regime
     since the Hubbard-I solution tends to overemphasize correlation and
     hence is quite unreliable at low values of U.
     The phase diagram shows
     that the density of electrons below which ferromagnetism can
     occur increases as $\tpr$/t is reduced below zero,
      \begin{center}
      \epsfig{file=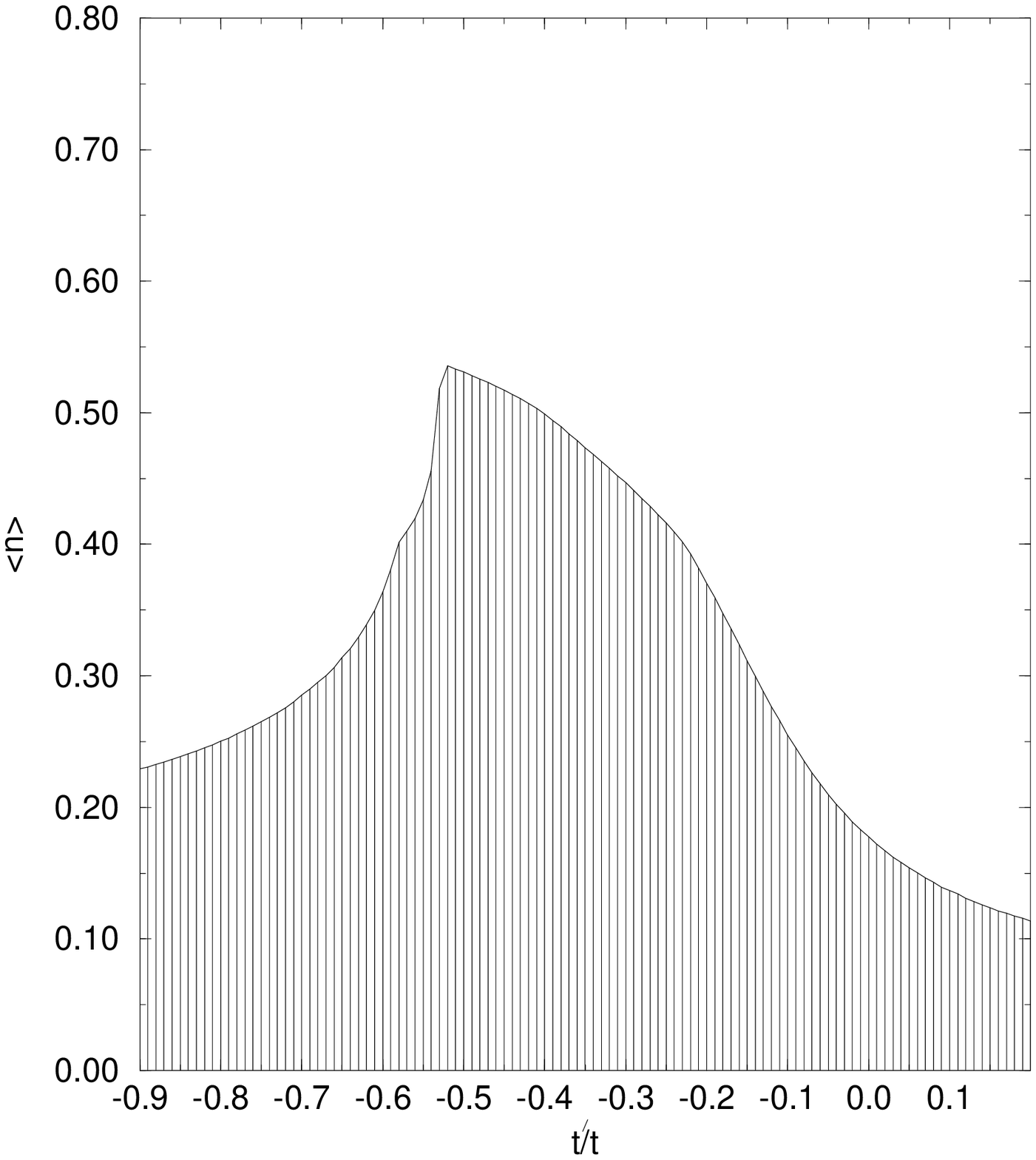,width=9cm,height=6cm}
      \vspace{-1.0cm}
      \begin{figure}[p]
      \caption{Phase diagram in the $\tpr$/t-$<n>$ plane for $U$=10. The
      shaded region denotes the ferromagnetic regime.
      \label{phase1}}
      \end{figure}
      \end{center}
     \begin{center}
     \epsfig{file=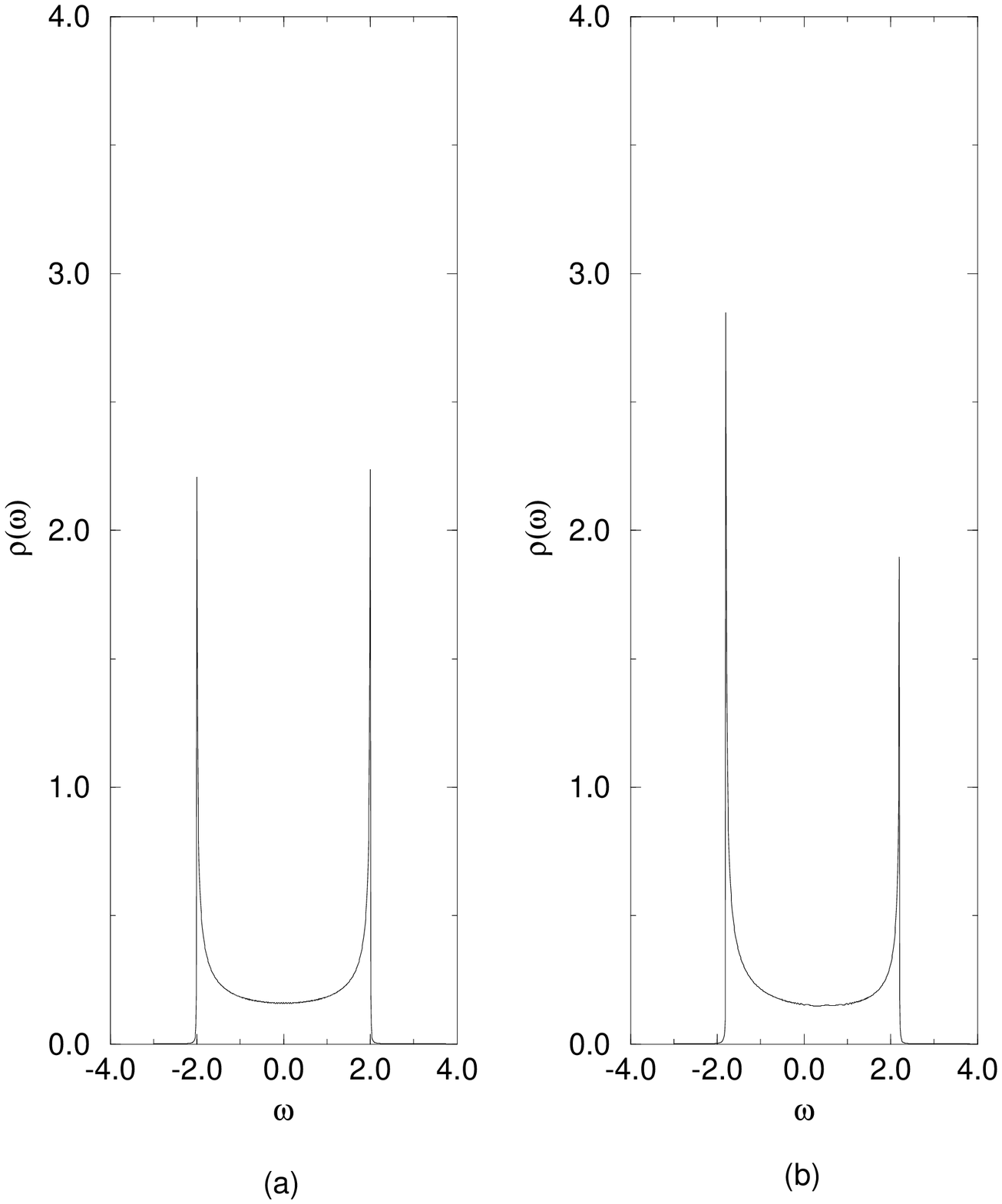,width=9cm,height=4.5cm}
     \epsfig{file=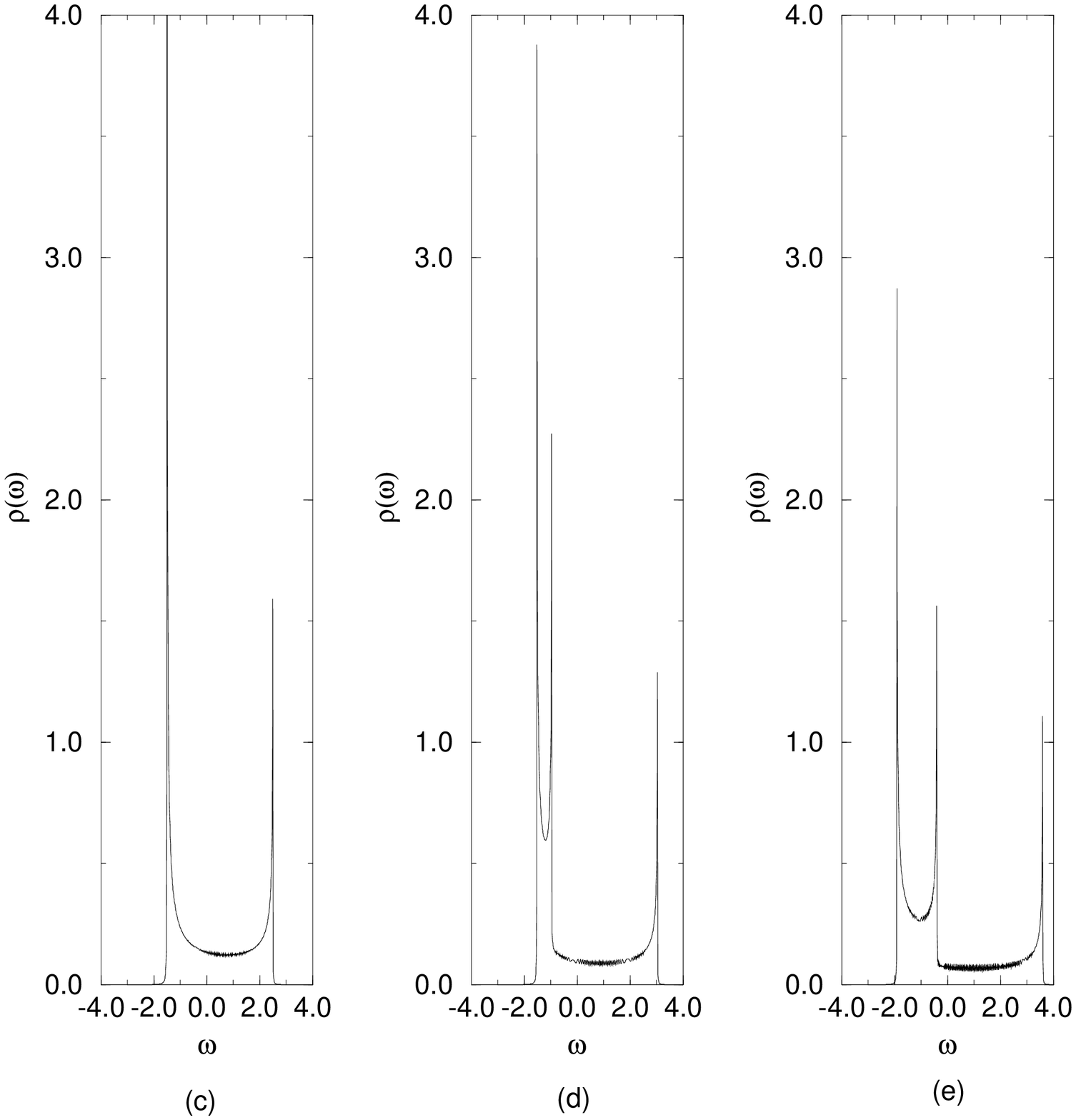,width=14cm,height=4.5cm}
     \vspace{-0.1cm}
     \begin{figure}[p]
      \caption{Density of states  
      for $U$=10 and $\tpr$/t = (a) 0.0, (b)
      -0.1, (c) -0.25, (d) -0.52, (e) -0.8. The figure shows only the
      Lower Hubbard Band since it is the only band which is relevant
      to our discussion. The free electron density of states looks 
      very similar.
      \label{dos}}
      \end{figure}
      \end{center}
     passes through a 
     maximum at $\tpr$/t $\approx$ -0.52 and then starts
     decreasing. 
     In order to understand the phase diagrams shown in Figs 2 and 3
     we have studied the behaviour of the density of states as a
     function of $\tpr/t$.
     The explanation for ferromagnetism in the t-$\tpr$-$U$ Model lies
     in the fact that the lower edge peak in the density of states is
     far away from the center of the band. This gives rise to
     ferromagnetism as argued in Hubbard-I. As
     $\tpr /t $ is reduced from zero, initially up to $\tpr/t$ =-0.25,
     this peak grows in height (Fig. 4), thus giving rise
     to larger and larger densities at which ferromagnetism can
     occur. 
     As $\tpr /t$ reduces towards -0.52, the height of the
     lower peak starts reducing again but it begins to develop a
     spread, which can accommodate larger density of electrons.
     Beyond $\tpr /t \approx$ -0.52, this spread begins to move
     towards the center of the band and hence, the critical density
     $<n>_c$ is reduced. 
     Thus, at $\tpr/t \approx$ -0.52, the area within the lower peak
     has an optimum value, thus giving a maximum value of $<n>_c$.

     In conclusion, the critical density $<n>_c$ below
     which ferromagnetism is possible in the t-$\tpr$-$U$ Model is
     governed by the density of states argument given in Hubbard-I.
     There are, however, some issues, which must be discussed here.
     As mentioned earlier, the Lieb-Mattis Theorem states that there
     can be no ferromagnetism for $\tpr = 0 $ ( which corresponds to
     the usual Hubbard Model) in one dimension. However, we find that
     within the Hubbard-I approximation, ferromagnetism occurs at
     $\tpr /t=0 $, at very low densities. This is essentially because
     of the approximation involved, whereby, although correlation is 
     taken into account beyond the Hatree Fock approximation, the solution,
     in essence, remains a mean field solution. However, in spite
     of the mean field nature of the phase diagram at $\tpr/t \neq 0 $
     gives an indication of the trend of the actual phase diagram
     which can be obtained by more accurate and exact calculations,
     work on which is in progress.


\end{document}